\documentclass[graybox]{svmult}

\usepackage{mathptmx}       
\usepackage{helvet}         
\usepackage{courier}        
\usepackage{type1cm}        
\usepackage{makeidx}         
\usepackage{graphicx}        
\usepackage{multicol}        
\usepackage[bottom]{footmisc}

\usepackage {amsmath, amssymb, mathscinet, color, xfrac}

\newtheorem{thm}{Theorem}[section]

\newtheorem{lem}[thm]{Lemma}
\newtheorem{defn}[thm]{Definition}

\numberwithin{thm}{section}
\newtheorem{lem/defn}[thm]{Lemma/Definition}
\newtheorem{preex/defn}[thm]{Example/Definition}
\newenvironment{ex/defn}%
  {\begin{preex/defn}\upshape}{\end{preex/defn}}

\newcommand{\ten}{\otimes}

\newcommand{\abs}[1]{\lvert#1\rvert}

\DeclareMathOperator{\End}{End}

\begin{document}

\title*{Boson-fermion correspondence of type B and twisted vertex algebras}
\author{Iana I. Anguelova}
\institute{Iana I. Anguelova \at College of Charleston, Math Department \\ 66 George Street \\ Charleston SC 29414, \ USA  \\ \email{anguelovai@cofc.edu}}
\maketitle
\abstract{
The  boson-fermion correspondence of type A  is an isomorphism between two super vertex algebras (and so has singularities in the operator product expansions only at  $z = w$). The boson-fermion correspondence of type B plays  similarly important role in many areas, including representation theory, integrable systems, random matrix theory and random processes. But the vertex operators describing it have singularities in their operator product expansions
at both $z = w$ and $z = -w$, and thus need a more general notion than that of a super vertex algebra.  In this paper we present such a notion: the concept of a twisted vertex algebra, which generalizes the concept of super vertex algebra.  The two sides of the  correspondence of type B constitute  two examples of twisted vertex algebras. The boson-fermion correspondence of type B is thus  an isomorphism between two twisted vertex algebras.
}

\section{Introduction}
\label{sec:intro}

In 1+1 dimensions (1 time and 1 space dimension) the bosons and fermions  are related by the boson-fermion correspondences. The simplest, and best known,  case of a boson-fermion correspondence is that of type A, but there are other examples of boson-fermion correspondences, for instance the boson-fermion correspondence of type B, the super boson-fermion correspondences of type A and B, and others.  They are extensively studied in many physics and mathematics papers, some of the first and most influential being the papers  by Date-Jimbo-Kashiwara-Miwa,  Igor Frenkel, Sato and Segal-Wilson, which make the connection between the representation theory of Lie algebras and soliton theory.  (Exposition of some of the mathematical results concerning the boson-fermion correspondence of type A are given in \cite{KacRaina},  \cite{Miwa-book}.) As with any mathematical concept, there are at least two distinct directions of inquiry. One is: what types of applications and structures we can get as a result of such a boson-fermion correspondence.  And the second direction addresses the fundamental questions: What \textbf{is} a boson-fermion correspondence? A correspondence of what mathematical structures?
For the simplest boson-fermion correspondence, that  of type A (often called the charged free boson-fermion correspondence), both these directions of inquiry have been addressed, and applications thereof continue to be found.
The first of these directions   was also  studied first, and the structures, properties, and applications of this  boson-fermion correspondence turned out to be very rich and many varied. As was mentioned above, Date-Jimbo-Kashiwara-Miwa and Igor Frenkel  discovered its connection to the theory of integrable systems, namely to the KP and KdV hierarchies, to the theory of symmetric polynomials and representation theory of infinite-dimensional Lie algebras, namely the $a_{\infty}$ algebra,  (whence the name "type A" derives), as well as  to the  $\hat{sl}_n$ and other \ affine Lie algebras. Their work sparked further interest in it, and there are now connections to many other areas, including number theory  and  geometry, as well as random matrix theory and random processes (see for example papers by Harnad, Orlov and van de Leur). As this boson-fermion correspondence turned out to have so many applications and connections to various mathematical areas, the natural question needed to be addressed: what \textbf{is} a boson-fermion correspondence--a correspondence of what mathematical structures?
A partial answer early on was given by Igor Frenkel in \cite{Frenkel-BF}, but the full answer \ had to wait for the development of the theory of vertex algebras. Vertex operators were introduced in the earliest days of string theory and now play an important role in many areas such as quantum field theory, integrable models, representation theory, random matrix theory,  and many others. The theory of super vertex algebras axiomatizes the properties of some, simplest, "algebras" of vertex operators (see for instance \cite{BorcVA}, \cite{FLM}, \cite{FHL}, \cite{Kac}, \cite{LiLep}). Thus, the answer to the question "what \textbf{is} the boson-fermion correspondence of type A"  is: the boson-fermion correspondence of type A is an isomorphism between two super vertex algebras (\cite{Kac}).

For other well-known boson-fermion correspondences, e.g. the type B, the super correspondence of type B, and others,  the question of applications and connections to other mathematical structures already has many answers. For example, Date, Jimbo, Kashiwara and Miwa, who introduced the correspondence of type B, discovered its connection to the theory of integrable systems, namely to the BKP hierarchy (\cite{DJKM-4}), to the representation theory of the $b_{\infty}$ algebra (whence the name "type B" derives), to symmetric polynomials and the symmetric group (some further developments were provided by You in \cite{YouBKP}). There are currently studies of its connection to random matrices and random processes by J. Harnad,  Van de Leur, Orlov,  and others (see for example \cite{OrlovVanDeLeur}).

On the other hand the question of "what \textbf{is} the boson-fermion correspondence of type B" has not been answered. We know it \textbf{is not} an isomorphism between two \textbf{super} vertex algebras anymore, as the correspondence of type A was. In this paper we  answer this question. To do that we need to  introduce the concept of a \textbf{twisted vertex algebra}, which generalizes the concept of super vertex algebra. The boson-fermion correspondence of type B is then   an isomorphism between two \textbf{twisted} vertex algebras.

The overview of the paper: first we briefly describe  the examples of the two super vertex algebras that constitute the boson-fermion correspondence of type A. We list only one property which is  the "imprint" of the boson-fermion correspondence of type A: namely the Cauchy determinant  identity that follows from the   equality between the vacuum expectation values of the two sides of the correspondence.  (As discussed above, there are many, many properties and applications of  any boson-fermion correspondence,  which can, and do,  occupy many papers).

Next we proceed with the definition of a twisted vertex algebra and the examples of the two twisted vertex algebras that constitute the boson-fermion correspondence of type B. Then again we list only one property which is the "imprint" of the boson-fermion correspondence of type B: namely the Schur Pfaffian  identity that follows from the   equality between the vacuum expectation values of the two sides of the correspondence.

\section{Super vertex algebras and boson-fermion correspondence of type A}
\label{sec:type A}

The following definition is well known, it can be found for instance in \cite{FLM}, \cite{FHL},  \cite{Kac}, \cite{LiLep} and others.  We recall it for completeness, as "algebras" of fields are the subject of this paper. (Roughly speaking, all vertex algebras, be they super or twisted,  are "singular algebras" of fields).
\begin{defn}
\begin{bf} (Field)\end{bf}
 A field $a(z)$ on a vector space $V$ is a series of the form
\begin{displaymath}
a(z)=\sum_{n\in \mathbf{Z}}a_{(n)}z^{-n-1}, \ \ \ a_{(n)}\in
\End(V),\ \ \text{such that }\ a_{(n)}v=0 \ \ \text{for any}\ v\in V, \ n\gg 0.
\end{displaymath}
\end{defn}
We also recall the following notation:
For a rational function $f(z,w)$ we denote by $i_{z,w}f(z,w)$
the expansion of $f(z,w)$ in the region $\abs{z}\gg \abs{w}$, and correspondingly for
$i_{w,z}f(z,w)$. Similarly, we will denote by $i_{z_1, z_2,\dots, z_n}$ the expansion in the region \mbox{$|z_1|\gg \dots \gg|z_n|$}. And lastly, we work with the category of super vector spaces, i.e., $\mathbb  {Z}_{2}$ graded vector spaces. The flip map $\tau $  is defined by
\begin{equation}
\label{eq:flip}
 \tau (a\ten b) =(-1)^{\tilde{a} \cdot \tilde{b}} (b\ten a)
\end{equation}
for any homogeneous elements $a, b$ in the super vector space, where $\tilde{a}$, $\tilde{b}$ denote correspondingly  the parity of $a$, $b$.

The definition of a super vertex algebra is well known, we refer the reader for example to  \cite{FLM}, \cite{FHL}, \cite{Kac}, \cite{LiLep}, as well as for  \ notations, details and theorems. We only  remark that (classical) vertex algebras have two  important properties which we would like to carry over to
the case of twisted vertex algebras. These are the analytic continuation and completeness with respect to  Operator Product Expansions (OPEs). In fact our definition of a twisted vertex algebra is based on enforcing these two properties. Recall we have for the OPE of two fields
\begin{equation}
a(z)b(w)=\sum _{j=0}^{N-1} i_{z,w}\frac{c^j(w)}{(z-w)^{j+1}} +:a(z)b(w):,
\end{equation}
where $:a(z)b(w):$ denotes the nonsingular part of the expansion of
$a(z)b(w)$ as a Laurent series in $(z-w)$.  We call  $:a(z)b(z):$ a
\emph{normal ordered product} of the fields $a(z)$ and $b(z)$.
 Moreover, $Res_{z=w}a(z)b(w)(z-w)^j=c^j(w)=(a_{(j)}b) (w)$, i.e., the coefficients of the OPEs are fields in the \textbf{same} super vertex algebra. Since for the commutation relations only the singular part of the OPEs matters, we abbreviate the OPE above as:
\begin{equation}
\label{eqn:OPE}
a(z)b(w)\sim \sum _{j=0}^{N-1}\frac{c^j(w)}{(z-w)^{j+1}}.
\end{equation}

For many examples, super vertex algebras are generated by a much smaller number of generating  fields,  with imposing the condition that the resulting space  of fields of the vertex algebra has to be closed under certain operations: For any field $a(z)$ the field $Da(z) =\partial_z a(z)$  has to be a field in the vertex algebra. Also, the  OPEs coefficients  ($c^j(w)$ from  \eqref{eqn:OPE})  and  normal ordered  products $:a(z)b(z):$ of  any two fields $a(z)$ and $b(w)$ have to be  fields in  the vertex algebra.
 Note that the identity operator on $V$ is always a trivial field in the vertex algebra, corresponding to the vacuum vector $|0\rangle  \in V$. The OPEs  are a good indicator of the restrictions placed by the definition of the super vertex algebra: for \ example, the only functions allowed in the OPEs when the identity field is the coefficient are the $\frac{1}{(z-w)^j}$ with $j\in \mathbb{N}$ (this is clearly seen in \eqref{equation:OPE-A}). We will use this information later as a way to compare the different generalizations of vertex algebras existing in the math literature, see Remark \ref{remark:OPEs} later.

Let us thus turn  our attention to the boson-fermion correspondence of type A.
The fermion side of the boson-fermion correspondence of type A is a super vertex algebra generated by  two nontrivial odd fields---two charged fermions: the fields $\phi (z)$ and $\psi (z)$ with only nontrivial operator product expansion (OPE) (see e.g. \cite{Miwa-book}, \cite{KacRaina} and \cite{Kac}):
\begin{equation}
\label{equation:OPE-A}
\phi (z)\psi (w)\sim \frac{1}{z-w}\sim \psi(z)\phi(w),
\end{equation}
where the $1$ above denotes the identity map $Id$.
The modes $\phi_n$ and $\psi_n$, $n\in \mathbf{Z}$ of the fields  $\phi (z)$ and $\psi (z)$, which we index as follows:
\begin{equation}
\phi (z) =\sum _{n\in \mathbf{Z}} \phi_n z^{n}, \quad \psi (z) =\sum _{n\in \mathbf{Z}} \psi_n z^{n},
\end{equation}
form a Clifford algebra $\mathit{Cl_A}$  with relations
\begin{equation}
[\phi_m,\psi_n]_{\dag}=\delta _{m+n, -1}1, \quad [\phi_m,\phi_n]_{\dag}=[\psi_m,\psi_n]_{\dag}=0.
\end{equation}
The indexing of the generating fields vary depending on the point of view; our indexing here corresponds to $\phi_n =\hat{v}_{n+1}, \quad \psi_n=\check{v}^*_{-n}$ in \cite{KacRaina}. This indexing and the properties of the vertex algebra dictate that
the underlying  space of states of this super vertex algebra---the fermionic Fock space --- is the highest weight   representation of $\mathit{Cl_A}$ generated by the vacuum  vector $|0\rangle $,  so that $\phi_n|0\rangle=\psi_n|0\rangle=0 \ \text{for} \  n<0$. \\
We denote both the space of states and the resulting vertex algebra generated by the fields $\phi (z)$ and $\psi (z)$ by  $\mathit{F_A}$. It is often called  the charged free fermion vertex algebra.

We can calculate vacuum expectation values if we have a  symmetric bilinear form $\langle
\ \mid\ \rangle: V\ten V \to \mathbb{C}$ on the space of states of the vertex algebra $V$.  Recall\footnote{There is a very important concept  of an invariant bilinear form on a vertex algebra, for details see for example \cite{FHL}, \cite{Li-bilinear}.} it is required that  the bilinear form is such that the vacuum vector $1=|0\rangle$ spans an  orthogonal subspace on its own, and that the bilinear form is  normalized on the vacuum vector.
By abuse of notation we will just write $\langle  0 \mid 0 \rangle$ instead of $\langle \langle 0| \mid |0\rangle \rangle$.
\begin{lem}
The following determinant formula for the vacuum expectation values on the fermionic side $\mathit{F_A}$ holds (\cite{Miwa-book}):
\begin{equation*}
 \langle  0 |\phi (z_1)\phi (z_2)\dots \phi (z_n)\psi (w_1)\psi (w_2)\dots \psi (w_n) | 0 \rangle =(-1)^{n(n-1)/2} i_{z;w} det \Big(\frac{1}{z_i -w_j}\Big)_{i, j =1}^n.
\end{equation*}
Here $i_{z; w}$  stands for the expansion $i_{z_1,z_2,\dots,z_n, w_1, \dots, w_n}$.
\end{lem}
The proof is usually given using Wick's formula, see \cite{Miwa-book}, although in \cite{AngTwisted} we give a proof depending entirely on  the underlying Hopf algebra structure similar to Laplace pairing.

The boson-fermion correspondence of type A is  determined once we write the images of  generating fields $\phi (z)$ and $\psi (z)$ under the correspondence. In order to do that, an \textbf{essential} ingredient is the so-called Heisenberg field $h(z)$ given by
\begin{equation}
h(z)=:\phi (z)\psi (z):
\end{equation}
It follows that the Heisenberg field $h(z)=\sum _{n\in \mathbb{Z}} h_n z^{-n-1}$ has OPEs with itself given by:
\begin{equation}
h(z)h(w)\sim \frac{1}{(z-w)^2}, \quad \text{in \ modes:} \ [h_m,h_n]=m\delta _{m+n,0}1.
\end{equation}
i.e., its  modes $h_n, \ n\in \mathbb{Z}$, generate a Heisenberg algebra $\mathcal{H}_{\mathbb{Z}}$.
It is well known that any irreducible highest weight module of this Heisenberg algebra is isomorphic to the polynomial algebra with infinitely many variables $\mathit{B_m}\cong \mathbb{C}[x_1, x_2, \dots , x_n, \dots ]$.
The fermionic Fock space decomposes (via the charge decomposition, for details see for example \cite{KacRaina}) as $\mathit{F_A}=\oplus _{i\in \mathbb{Z}} B_i$, which we can write as
\begin{equation}
\mathit{F_A}=\oplus _{i\in \mathbb{Z}} B_i \cong \mathbb{C}[e^{\alpha}, e^{-\alpha}]\ten \mathbb{C}[x_1, x_2, \dots , x_n, \dots ],
\end{equation}
where by $\mathbb{C}[e^{\alpha}, e^{-\alpha}]$ we mean the Laurent polynomials with one variable   $e^{\alpha}$. \begin{footnote}{The reason for this notation is that the resulting vertex algebra is a lattice vertex algebra.}\end{footnote} The isomorphism is as  Heisenberg modules, where $e^{n\alpha}$ is identified as the highest weight vector for the irreducible Heisenberg module $B_n$. We denote the  vector space on the right-hand-side of this $\mathcal{H}_{\mathbb{Z}}$-module isomorphism by $B_A$.  $B_A$ is then the underlying vector space of the bosonic side of the boson-fermion correspondence of type A.

Now we can write the images of  generating fields $\phi (z)$ and $\psi (z)$ under the correspondence:
\begin{equation}
\phi (z)\mapsto e^{\alpha}(z), \quad  \psi (z)\mapsto e^{-\alpha}(z),
\end{equation}
where the generating fields $e^{\alpha}(z)$, $e^{-\alpha}(z)$ for the bosonic part of the correspondence are given by
\begin{equation*}
e^{\alpha}(z)=\exp (\sum _{n\ge 1}\frac{h_{-n}}{n} z^n)\exp (-\sum _{n\ge 1}\frac{h_{n}}{n} z^{-n})e^{\alpha}z^{\partial_{\alpha}},
\end{equation*}
\begin{equation*}
e^{-\alpha}(z)=\exp (-\sum _{n\ge 1}\frac{h_{-n}}{n} z^n)\exp (\sum _{n\ge 1}\frac{h_{n}}{n} z^{-n})e^{-\alpha}z^{-\partial_{\alpha}},
\end{equation*}
the operators $e^{\alpha}$, $e^{-\alpha}$, $z^{\partial_{\alpha}}$ and $z^{-\partial_{\alpha}}$ act in an obvious way on the space $B_A$.

The resulting super vertex algebra generated by the fields $e^{\alpha}(z)$ and $e^{-\alpha}(z)$ with underlying vector space $B_A$ we denote also by $B_A$.
\begin{lem}
The following product formula for the vacuum expectation values on the bosonic side $B_A$ holds:
\begin{equation*}
\langle  0 |e^{\alpha} (z_1)e^{\alpha} (z_2)\dots e^{\alpha} (z_n)e^{-\alpha} (w_1)e^{-\alpha} (w_2)\dots e^{-\alpha} (w_n) | 0 \rangle =i_{z; w} \frac{\prod_{i<j}^n ((z_i-z_j)(w_i-w_j))}{\prod_{i, j=1}^n (z_i-w_j)}
\end{equation*}
Here $i_{z; w}$  stands for the expansion $i_{z_1,z_2,\dots,z_n, w_1, \dots, w_n}$.
\end{lem}
\begin{thm}(\cite{Kac})
The boson-fermion correspondence of type A is the isomorphism between the charged free fermion super vertex algebra $\mathit{F_A}$ and the bosonic super vertex algebra $B_A$.
\end{thm}
\begin{lem} The Cauchy's determinant identity follows from the equality of the vacuum expectation values:
\begin{align*}
(-1)^{n(n-1)/2} det \Big(\frac{1}{z_i -w_j}\Big)_{i, j =1}^n =AC \langle  0 |\phi (z_1)\dots \phi (z_n)\psi (w_1)\dots \psi (w_n) | 0 \rangle =\\
=AC \langle  0 |e^{\alpha} (z_1)\dots e^{\alpha} (z_n)e^{-\alpha} (w_1)\dots e^{-\alpha} (w_n) | 0 \rangle =\frac{\prod_{i<j} (z_i-z_j)\prod_{i<j} (w_i-w_j)}{\prod_{i, j=1}^{n} (z_i-w_j)}
\end{align*}
$AC$ stands for Analytic Continuation.
\end{lem}
 The Cauchy determinant identity (the equality between  the first and the fourth expressions) is a historic identity and is well known, one of the oldest references being \cite{littlewood}. The proof of it is   usually given using factorization, remarkably even in \cite{Miwa-book}, see Remark 5.1 there. The point here is that the Cauchy identity follows immediately from the equality of the vacuum expectation values of both sides of the boson-fermion correspondence, and is a quintessential ``imprint" of the correspondence (although  this identity is absent in some standard mathematical references of the boson-fermion correspondence of type A like \cite{KacRaina} and \cite{Kac}).

\section{Twisted vertex algebras and boson-fermion correspondence of type B}
\label{sec:type B}

Here we will only give the definition for a twisted vertex algebra of order 2, for more general definition and details see \cite{AngTwisted}. We begin with some preliminaries.
\begin{defn}
The Hopf algebra $H_{T_{-1}}$ is the Hopf algebra with  a primitive generator $D$ and a grouplike generator $T_{-1}$ subject to the  relations:
\begin{equation}
DT_{-1 }=-T_{-1 } D, \quad \text{and} \ (T_{-1 })^2=1
\end{equation}
\end{defn}
Denote by $\mathbf{F}^2_{\pm}(z, w)$ the space of rational functions in the variables $z, w\in \mathbb{C}$ with only poles at $z=0, \ z= \pm w$. Note that we do not allow poles at $w=0$, i.e., if $f(z, w)\in \mathbf{F}^2_{\pm}(z, w)$, then $f(z, 0)$ is well defined.
Similarly, $\mathbf{F}^2_{\pm}(z_1, z_2, \dots , z_l)$ is the space of rational functions in variables $z_1, z_2, \dots, z_l$ with only poles at $z_1=0$, or $z_j= \pm z_k$.
$\mathbf{F}^2_{\pm}(z, w)$ is a  $H_{T_{-1}}\ten H_{T_{-1}}$  Hopf algebra module by
\begin{eqnarray}
&D_z f(z, w)=\partial_z f(z, w), \quad (T_{-1}) _{z} f(z, w)=f(-z, w) \\
&D_w f(z, w)=\partial_w f(z, w), \quad (T_{-1}) _{w} f(z, w)=f(z, -w)
\end{eqnarray}
We will denote the action of elements $h\ten 1 \in H_{T_{-1}}\ten H_{T_{-1}}$  on $\mathbf{F}_{\pm}(z, w)$ by $h_z\cdot$, and similarly $h_w\cdot$ will denote the action of elements $1\ten h \in H_{T_{-1}}\ten H_{T_{-1}}$.
\begin{defn}\label{defn:twistedVA} \begin{bf}(Twisted vertex algebra of order $2$)\end{bf}\\
Twisted vertex algebra of order $2$ is a collection of  the following data:
\begin{itemize}
\item the space of states: a vector space $W$;
\item the space of fields: a vector super space $V$-- an $H_{T_{-1}}$ module, such that $V\supset W$;
\item a projection: a linear map $\pi_{f}:V\to W$;
\item a field-state correspondence: a linear map  from $V$ to the space of fields on $W$;
\item a vacuum vector: a vector $1=|0\rangle \in W\subset V$.
\end{itemize}
This data should satisfy the following  set of axioms:
\begin{itemize}
\item Vacuum axiom: \ \ $Y(1, z)=Id_W$;
\item  Modified creation axiom: \ \ $Y(a, z)1\arrowvert _{z=0}=\pi_f(a)$, for any $a\in V$;
\item Transfer of action:\ \  $Y(ha,z)=h_z\cdot Y(a, z)$ for any $h\in H_{T_{-1}}$;
\item Analytic continuation:  For any $a_i \in V, i=1,\dots ,k$,  the composition  \\
\mbox{$Y(a_1,z_1)Y(a_2, z_2)\dots Y (a_k, z_k)1$} converges in the domain  \mbox{$|z_1|\gg \dots \gg|z_k|$} and can be continued to a  rational vector valued   function
    \begin{equation*}
    X_{z_1, z_2, \dots , z_k}(a_1\ten a_2\ten \dots \ten a_k):V^{\ten k} \to W\ten \mathbf{F}^2_{\pm}(z_1, z_2, \dots , z_k),
    \end{equation*}
    so that $Y(a_1, z_1)Y(a_2, z_2)\dots Y (a_k, z_k)1=i_{z_1, z_2,\dots,z_k} X_{z_1, z_2, \dots , z_k}(a_1\ten a_2\ten \dots \ten a_k)$
\item Supercommutativity: $X_{z, w}(a\ten b) =X_{w, z}(\tau (a\ten b))$, with $\tau$ defined in \eqref{eq:flip}.
\item Completeness with respect to OPEs (modified): For each $n\in \mathbb{N}$ there exists $l_n\in \mathbf{Z}$ such that
$Res_{z=\pm  w}X_{z, w, 0}(a\ten b\ten v)(z\mp w)^n =Y(c_n,w)\pi_f(v)w^{l_n} \ \ \text{for some} \ \ c_n\in V.$
\end{itemize}
\end{defn}
\begin{remark}
The axiom/property  requiring completeness with respect to the OPEs is a weaker one than in the classical vertex algebra case. We can express this weaker axiom by saying that the modes of the OPE coefficients, the residues $Res_{z=\pm w}(z\mp w)^nY(a, z)Y(b, w)$, are the  modes of a field that belongs to the twisted algebra, modulo a shift (a shift by $w^{l_n}$ is allowed, but no more). The stronger property is violated in the interesting examples, see for instance Remark \ref{remark:OPEs} below. \qed
\end{remark}
\begin{remark}
The axiom of analytic continuation expresses two requirements: one,  that the products of fields are expansions  of  \textbf{rational} operator valued functions in appropriate regions; and two, that the only poles of these rational functions are  at $z=\pm w$. Thus, if we think of the variable $z$ as being a square root of another variable $\tilde{z}$ (similarly $w=\sqrt{\tilde{w}}$), then the only allowed singularities are at $z^2=w^2$, i.e., at $\tilde{z}=\tilde{w}$, which is a prerequisite for the usual locality axiom (in the variable $\tilde{z}$). The usual  locality axiom  requires not only that the singularities are located only at  $\tilde{z}=\tilde{w}$, but also that the supercommutativity axiom $X_{z, w}(a\ten b) =X_{w, z}(\tau (a\ten b))$ holds. These two axioms combined produce the commutation or anticommutation relations obeyed by  the bosons or fermions correspondingly. If we remove the supercommutativity axiom, then we can  have a more general braided locality, instead of the usual locality, and there are examples (e.g., the quantum affine Lie algebras at roots of unity) which do not obey the supercommutativity axiom. But  the examples of boson-fermion correspondences do indeed obey the  supercommutativity axiom, which is why we have required it as part of our definition of a twisted vertex algebra. \qed
\end{remark}
\begin{remark}
There are other generalizations of the notion of super vertex algebra, for instance there is the notion(s) of a twisted module for a vertex algebra (see \cite{FLM}, \cite{BakKac}); there are also the notions of a generalized vertex algebra (see e.g. \cite{DongLepVA}, \cite{KacGenVA}). The notion of a twisted vertex algebra as outlined above is different from any of those notions in two main respects: the first difference is in the functions allowed in the OPEs (see Remark \ref{remark:OPEs} below); the second is the fact that the space of fields is larger than the space of states (the space of states in the examples here is a proper projection of the space of fields). In this, the notion of twisted vertex algebra resembles the notion of a Deformed Chiral Algebra of \cite{FR}, and although there are  differences, one can think of a twisted vertex algebra as being the root of unity symmetric version of the Deformed Chiral Algebra concept.
\qed
\end{remark}
 Similarly to super vertex algebras,  twisted vertex algebras are often generated by a smaller number of fields (for a rigorous theorem regarding that see \cite{AngTwisted}). The space of  fields is then  determined  by  requiring, as before,  that it is closed under OPEs (see modification above). Also, as before, for any field $a(z)$ the field $Da(z) =\partial_z a(z)$  again has to be a field in the twisted vertex algebra. But now we also require that the field $T_{-1}a(z)=a(-z)$ is  a field in the twisted vertex algebra of order 2 as well. Note that this immediately violates the  stronger creation axiom for a classical vertex algebra, hence any such field cannot belong to a classical vertex algebra. This is the reason we require the  modified  field-state correspondence with the \textbf{modified creation axiom} for a twisted vertex algebra.

 We now proceed with the two examples of a twisted vertex algebra of order 2 which give the two sides of the boson-fermion correspondence of type B. The fermionic side  is generated by a single field $\phi (z)=\sum _{n\in \mathbf{Z}} \phi_n z^{n}$, with OPEs with itself given by (\cite{DJKM-4}, \cite{YouBKP}--modulo a factor of 2; \cite{AngTwisted}):
\begin{equation}
\label{equation:OPE-B}
\phi (z)\phi (w)\sim \frac{z-w}{z+w}, \quad \text{in \ modes:} \ [\phi_m,\phi_n]_{\dag}=2(-1)^m\delta _{m, -n}1.
\end{equation}
Thus the  modes generate a Clifford algebra $\mathit{Cl_B}$, and the underlying space of states, denoted by $\mathit{F_B}$, of the twisted vertex algebra is a highest weight representation of $\mathit{Cl_B}$ with  the vacuum  vector $|0\rangle $, such that $\phi_n|0\rangle=0 \ \text{for} \  n<0$.   The space of fields, which is larger than the space of states,  is generated by the field $\phi (z)$ together with its descendent $ T_{-1}\phi (z)=\phi (-z)$. We call the resulting twisted vertex algebra the \textbf{free neutral fermion of type B}\begin{footnote}{The reason for the name is that there is a free neutral fermion of type D, which is  commonly referred to  as just "the free neutral fermion". In fact, there is a boson-fermion correspondence of type D-C, see \cite{AngTwisted}.}\end{footnote}, and denote also by $\mathit{F_B}$.
\begin{remark}
\label{remark:OPEs}
If we look at the defining OPE, \eqref{equation:OPE-B}, we can see that if we just write the singular part, we have
the residue
$Res_{z=-w}\phi (z)\phi (w)=-2w\cdot 1=-2wId_W$, which can not be a field in any vertex algebra as it is. But a shift by $w^{-1}$ will produce the field  $-2Id_W$, which is the field corresponding to the $-2|0\rangle$. This exemplifies that in the OPEs of twisted vertex algebras when  the identity field is the coefficient we do allow any function of the type $\frac{w^k}{(z\pm w)^l}$, where $k\in \mathbb{Z}_{\ge 0},\  l\in \mathbb{N}$. For instance, besides the classical vertex algebra singularity $\frac{1}{z-w}$, we allow  additionally, and on its own, $\frac{1}{z+w}$, which is not allowed in twisted modules for vertex algebras, nor in the notions of  generalized vertex algebras. In contrast, if the identity field is the coefficient, twisted modules for vertex algebras would only allow singularities in its OPEs of the type
\begin{equation*}
\partial_{\tilde{w}}^l \frac{\sqrt{\tilde{z}}}{\sqrt{\tilde{w}}} \frac{1}{\tilde{z}-\tilde{w}} =\partial_{w^2} ^l \frac{z}{w(z^2-w^2)}, \quad l\in \mathbb{Z}_{\ge 0},
 \end{equation*}
 i.e., even though the singularities for twisted modules are indeed at $z=\pm w$, only particular combinations of $\frac{1}{z-w}$, $\frac{1}{z+w}$ and derivatives are allowed (see e.g. \cite{BakKac}).
 In the case of  a generalized vertex algebra   the singularities allowed in the OPEs when the  vacuum field is the coefficient  are of the type $\frac{1}{(\tilde{z}-\tilde{w})^{\sfrac{1}{N}}}=\frac{1}{(z^2 -w^2)^{\sfrac{1}{N}}}, \ \ N\in \mathbb{N}$.
\qed
\end{remark}
The boson-fermion correspondence of type B is again determined once we write the image of the generating fields $\phi (z)$ (and thus of $T_{-1}\phi (z)=\phi(\small{-}z)$) under the correspondence. In order to do that, an essential ingredient is once again the  \textbf{twisted} Heisenberg field $h(z)$ given by\footnote{For \ details on normal ordered products in this more general case see \cite{AngTwisted}, the construction uses an additional Hopf algebra structure, similar to Laplace pairing.}
\begin{equation}
h(z)=\frac{1}{4}:\phi (z)T_{-1}\phi (z) := \frac{1}{4}:\phi (z)\phi (-z) :
\end{equation}
It follows that the twisted Heisenberg field, which due to  the symmetry above  has only  odd-indexed modes,   $h(z)=\sum _{n\in \mathbb{Z}} h_{2n+1} z^{-2n-1}$, has OPEs with itself given by:
\begin{equation}
h(z)h(w)\sim \frac{zw(z^2 +w^2)}{2(z^2 -w^2)^2},
\end{equation}
Its  modes, $h_n, \ n\in 2\mathbb{Z}+1$, generate a \textbf{twisted} Heisenberg algebra $\mathcal{H}_{\mathbb{Z}+\sfrac{1}{2}}$ with relations $[h_m,h_n]=\frac{m}{2}\delta _{m+n,0}1$, \ $m,n$ are now odd  integers.  It has (up-to isomorphism) only one irreducible module $B_{\sfrac{1}{2}}\cong \mathbb{C}[x_1, x_3, \dots , x_{2n+1}, \dots ]$.
The fermionic space of states $\mathit{F_B}$ decomposes as $\mathit{F_B} =B_{\sfrac{1}{2}} \oplus B_{\sfrac{1}{2}}$ (for \ details, see
\cite{DJKM-4}, \cite{YouBKP}, \cite{AngTwisted}). We can write this as an isomorphism of twisted Heisenberg modules for $\mathcal{H}_{\mathbb{Z}+\sfrac{1}{2}}$  in a similar way to the type A correspondence:
\begin{equation}
\mathit{F_B}=B_{\sfrac{1}{2}} \oplus B_{\sfrac{1}{2}} \cong \mathbb{C}[e^{\alpha}, e^{-\alpha}]\ten \mathbb{C}[x_1, x_3, \dots , x_{2n+1}, \dots ],
\end{equation}
but now we have  the extra relation $e^{2\alpha}\equiv 1$, i.e., $e^{\alpha}\equiv e^{-\alpha}$. The right-hand-side, which we denote by  $B_B$,  is the underlying vector space of \textbf{states} of the bosonic side of the boson-fermion correspondence of type B.

Now we can write the image of the  generating field $\phi (z)\mapsto e^{\alpha}(z)$, which will determine the correspondence of type B (for proof see \cite{AngTwisted}):
\begin{equation}
e^{\alpha}(z)=\exp \big(\sum _{k\ge 0}\frac{h_{-2k-1}}{k+\sfrac{1}{2}} z^{2k+1} \big)\exp \big(-\sum _{k\ge 0}\frac{h_{2k+1}}{k+\sfrac{1}{2}} z^{-2k-1} \big)e^{\alpha},
\end{equation}
The fields $e^{\alpha}(z)$ and  $e^{\alpha}(-z)=e^{-\alpha}(z)$ (observe the symmetry) generate the resulting \textbf{twisted} vertex algebra, which we denote also by $B_B$.

Note that one Heisenberg  $\mathcal{H}_{\mathbb{Z}+\sfrac{1}{2}}$-module $B_{\sfrac{1}{2}}$ on its own can be realized as a \textbf{twisted module} for \ an ordinary super vertex algebra (see \cite{FLM} for details), but the point is that we need \textbf{two} of them glued together for  the bosonic side of the correspondence. The two of them glued together as above no longer constitute a twisted module for an ordinary super vertex algebra.
\begin{thm}
The boson-fermion correspondence of type B is the isomorphism between the  fermionic \textbf{twisted} vertex algebra $\mathit{F_B}$ and the bosonic \textbf{twisted} vertex algebra $B_B$.
\end{thm}
\begin{lem} The Schur Pfaffian identity follows from the equality between the vacuum expectation values:
\begin{equation*}
AC \langle  0 |\phi (z_1)\dots \phi (z_{2n})| 0\rangle =Pf \Big(\frac{z_i -z_j}{z_i +z_j}\Big)_{i, j =1}^{2n} =\prod_{i<j}^{2n} \frac{ z_i-z_j}{z_i +z_j} =AC \langle  0 |e^{\alpha} (z_1)\dots e^{\alpha} (z_{2n}) | 0 \rangle
\end{equation*}
Here  $Pf$ denotes the Pfaffian of an antisymmetric matrix, $AC$ stands for Analytic Continuation..
\end{lem}
 Remark: the general definition of a twisted vertex algebra of order $N$, details and proofs can be found in \cite{AngTwisted}.

In conclusion,  we would like to thank the organizers of the International Workshop "Lie Theory and its Applications in Physics"  for a most enjoyable and productive workshop, and may it continue for many years to come!

\def\cprime{$'$}

\end{document}